\documentclass{webofc}

\usepackage[varg]{txfonts}
\usepackage{hyperref}
\usepackage{url}
\usepackage{amssymb}
\usepackage{lipsum}
\usepackage{amsmath}
\usepackage{graphicx}
\usepackage{esvect}
\usepackage{cancel}
\usepackage{physics}
\usepackage{float}
\usepackage{mathtools}
\usepackage{multirow}
\usepackage{braket}
\usepackage{gensymb}
\usepackage{natbib}

\hypersetup{colorlinks=true,citecolor=blue,urlcolor=blue,linkcolor=blue}
\begin{document}
\title{New Insights into Indirect Studies of the $\alpha(d,\gamma)\prescript{6}{}{\rm Li}$ \\
Radiative Capture}

\author{\firstname{Monica} \lastname{Sanjinez-Ortiz}\inst{1}\fnsep\thanks{\email{msanjine@uni-mainz.de}} \and
        \firstname{Pierre} \lastname{Capel}\inst{1}\fnsep\thanks{\email{pcapel@uni-mainz.de}}
}

\institute{Institut für Kernphysik, Johannes Gutenberg Universität-Mainz, Mainz, D-55099, Germany
          }

\abstract{The radiative capture reaction $\alpha(d,\gamma)\prescript{6}{}{\rm Li}$ plays an important role in the early-Universe production of lithium. Two attempts to determine its cross section at low energy indirectly via Coulomb dissociation were performed in the past. We report here on a new theoretical analysis of these Coulomb breakup experiments with a fully dynamical reaction model. Our results indicate that the breakup of $\prescript{6}{}{\rm Li}$ onto $\prescript{208}{}{\rm Pb}$ is nuclear dominated at forward angles as Coulomb breakup is suppressed due to the $\alpha$-d clustered structure of $\prescript{6}{}{\rm Li}$. We suggest that this suppression may be a general feature of breakup processes involving $N=Z$ clustered nuclei. We also identify the relevance of higher-order Coulomb effects in the breakup process, and conclude that extracting radiative capture cross sections from breakup data in this case is extremely challenging.
}
\maketitle
\section{Introduction}
\label{introduction}

The  $\alpha(d,\gamma)\prescript{6}{}{\rm Li}$ radiative capture reaction plays an important role in the picture of Big-Bang nucleosynthesis (BBN) for it is the main path for the production of primordial $\prescript{6}{}{\rm Li}$. The determination of its reaction rate is important for the assessment of isotopic lithium abundances determined by observation \cite{Spite_2009}, among others. However, direct measurements of its cross section remained prohibitive for several decades due to the challenges presented by low-energy measurement for reactions involving charged nuclei. 

The Coulomb breakup of $\prescript{6}{}{\rm Li}$ onto $\prescript{208}{}{\rm Pb}$ was used on two occasions as an indirect technique aiming to infer the astrophysical S-factor for this reaction. The first attempt was performed at the Karlsruhe Isochronous Cyclotron with a $\prescript{6}{}{\rm Li}$ beam at 26 MeV/nucleon \cite{Kiener}, and the second at the GSI in Darmstadt with a beam of 150 MeV/nucleon \cite{Hammache}. While the attempt at GSI did not lead to an extraction of the capture cross section from breakup data,  the S-factor extracted from the experiment at 26 MeV/nucleon has been in tension with theoretical predictions and the first direct low-energy measurements obtained by LUNA \cite{LUNA}. 

Here, we present a new theoretical reanalysis of these Coulomb breakup experiments to gain better insights into the discrepancy between direct and indirect measurements of the $\alpha(d,\gamma)\prescript{6}{}{\rm Li}$ reaction cross section as well as into the dynamics of breakup reactions \cite{sanjinez2026can}.

\section{Breakup-reaction analysis}

The analysis of the breakup reactions in this work was done within the Dynamical Eikonal Approximation (DEA) \cite{Baye}. The DEA has been extensively used to model the dissociation of weakly bound nuclei \cite{Goldstein}, and allows for the simultaneous inclusion of Coulomb and nuclear breakup effects into the reaction. In this analysis, $\prescript{6}{}{\rm Li}$ is described with the inter-fragment potential $V_{\alpha d}$ developed in Ref.~\cite{Hammache}. For the breakup reaction at 150 MeV/nucleon the cluster-target optical potentials were also taken from Ref. \cite{Hammache}, while for that at 26 MeV/nucleon, they were retrieved from global parametrizations at suitable energies \cite{Nolte, An}.

\subsection{Angular distribution of the breakup cross section}

\begin{figure*}[h]
\centering
\vspace*{0.1cm} 
\includegraphics[width=13cm,clip]{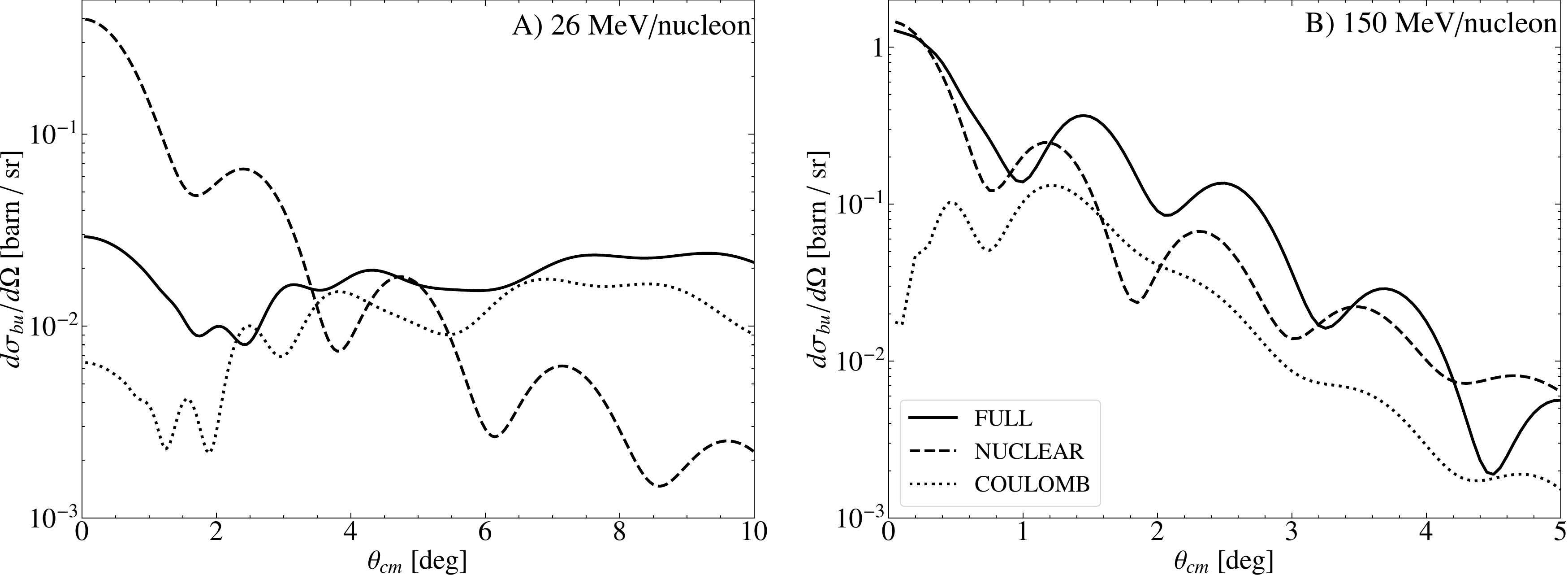}
\caption{Breakup cross section as a function of the scattering angle integrated over the energy range $E_{\alpha d}$ = $0$--$500 $ keV at A) 26 MeV/nucleon and B) 150 MeV/nucleon. Adapted from Ref.~\cite{sanjinez2026can}.}
\label{fig:angular}
\end{figure*}

Our results for the angular distribution of the breakup cross section integrated over the domain of relative energies relevant for BBN are shown in Fig.~\ref{fig:angular}. At each beam energy, three different calculations are shown: a full calculation, in which the $\prescript{6}{}{\rm Li}$ projectile and the $\prescript{208}{}{\rm Pb}$ target interact via both Coulomb and nuclear forces (solid lines), and also pure Coulomb (dotted lines) and pure nuclear (dashed lines) calculations. In the case of pure Coulomb breakup, a cut is imposed on the impact parameter at $b_{\rm min}=10$\,fm to simulate nuclear absorption.

We observe that at 150 MeV/nucleon, the full breakup process is strongly influenced by the nuclear contribution. This finding supports the original analysis of the experiment, which concluded that nuclear contributions were present in their measurements \cite{Hammache}. At 26 MeV/nucleon, we find that the Coulomb contribution plays a more prominent role. However, nuclear breakup remains very significant, being several times larger or comparable to the Coulomb contribution even at forward angles. 

We find that, regardless of the beam energy, the Coulomb contribution at forward angles in the dissociation of $\prescript{6}{}{\rm Li}$ onto $\prescript{208}{}{\rm Pb}$ is considerably suppressed with respect to semiclassical expectations resulting in a nuclear dominated process. In a semiclassical view, forward-angle contributions result from interactions at large impact parameters, where the Coulomb field of the target dominates. Nevertheless, in this reaction the $\alpha$ and deuteron fragments in the projectile are both $N=Z$ nuclei. Consequently, they possess approximately the same charge-to-mass ratio and, thus, are equally accelerated at large impact parameters by the Coulomb field resulting in no tidal force to induce the projectile's dissociation \cite{Baye_2018}. We then establish that such a suppression is an imprint of the $\alpha$-$d$ cluster structure of $\prescript{6}{}{\rm Li}$ onto the reaction dynamics \cite{sanjinez2026can}. We believe that this observation may have broader implications in the application of the Coulomb breakup method for the study of reactions involving other $N=Z$ clustered nuclei such as $\prescript{16}{}{\rm O}$ \cite{bertulani1994}.

\subsection{Triple differential breakup cross section}

\begin{figure}[h]
\centering
\sidecaption
\includegraphics[width=6.5cm,clip]{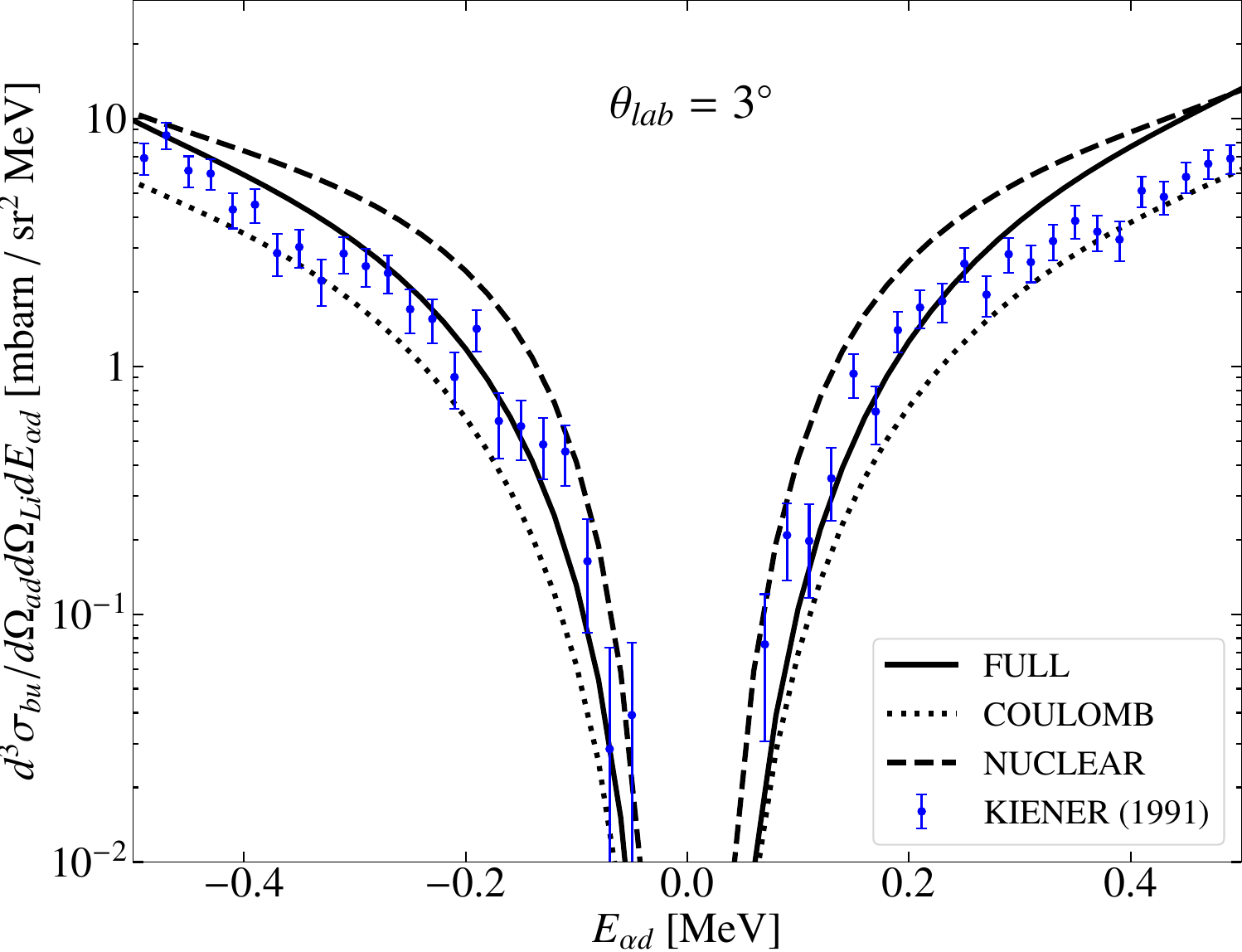}
\caption{Triple differential breakup cross section at $\theta_{lab} = 3 \degree$ compared to the data measured at 26 MeV/nucleon \cite{Kiener}. The positive and negative $E_{\alpha d}$ branches correspond to the forward and backward emission of the $\alpha$ in the reaction. Adapted from Ref.~\cite{sanjinez2026can}.}
\label{fig-2} 
\end{figure}

The experimental campaign at 26 MeV/nucleon collected measurements of the triple differential breakup cross section at $\theta_{lab} < 7 \degree$ assuming Coulomb breakup dominance \cite{Kiener}. Then, the astrophysical S-factor for the capture reaction was inferred from their data at $\theta_{lab} = 3 \degree$.  Our calculations for this breakup cross section at $\theta_{lab} = 3 \degree$ are presented in Fig.~\ref{fig-2} along with the corresponding data in the center-of-mass frame. We observe a good agreement between our full breakup prediction and the measurements, suggesting the S-factor inferred by this experiment was partially enhanced by additional nuclear breakup contributions.

\subsection{Partial-wave contributions to the breakup}

\begin{figure}[h]
\centering
\sidecaption
\includegraphics[width=6.5cm,clip]{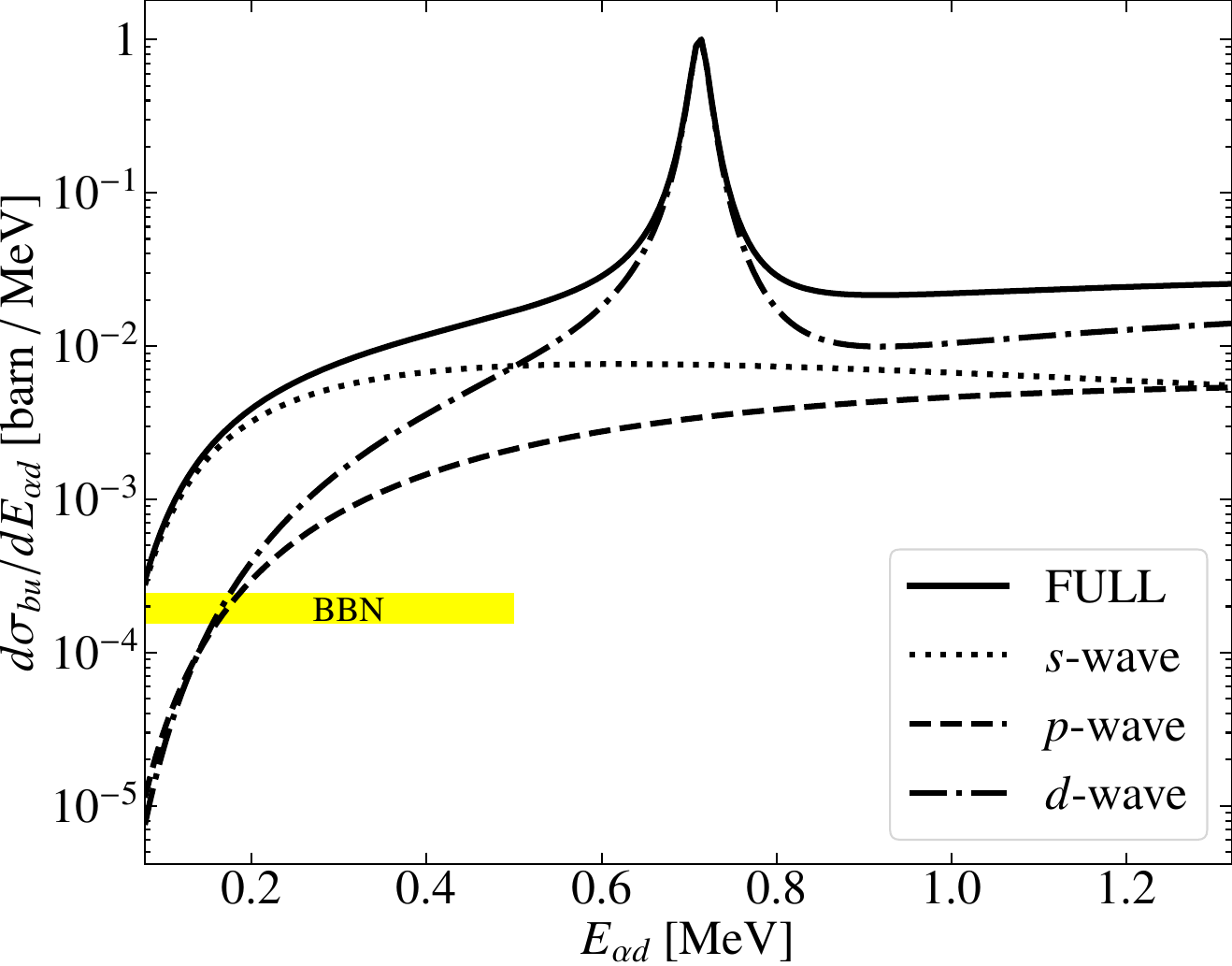}
\caption{Partial-wave decomposition of the full breakup cross section at 26 MeV/nucleon. From Ref.~\cite{sanjinez2026can}.}
\label{fig-3}
\end{figure}

The inference of the S-factor from the data at 26 MeV/nucleon treated the reaction at the first order of the perturbation theory and assumed that the breakup proceeded via pure electric quadrupole (E2) transitions \cite{Kiener}. For this reaction, this corresponds to transitions from the bound $s$- to continuum $d$-waves. To assess the validity of this assumption, we computed the partial wave decomposition of the breakup cross section, which is shown in Fig.~\ref{fig-3}. We observe that $s$- and $p$-waves on the continuum also have a relevant role. In the dynamical reaction model, the contributions of $s$- to $s$-wave transitions arise from nuclear breakup as well as higher-order contributions to the Coulomb process. These contributions, in particular, dominate the entire energy range of interest for BBN and, consequently, can lead to an unexpected inflation of the inferred astrophysical S-factor.

\section{Conclusions}

We have presented the results of a theoretical reanalysis \cite{sanjinez2026can} of the Coulomb breakup of $\prescript{6}{}{\rm Li}$ on $\prescript{208}{}{\rm Pb}$ experimentally determined in the past at 26 MeV/nucleon \cite{Kiener} and 150 MeV/nucleon \cite{Hammache}. Our findings confirm that at both beam energies the breakup of $\prescript{6}{}{\rm Li}$ is heavily influenced by the nuclear interaction with the $\prescript{208}{}{\rm Pb}$ target. In particular, we find that the process is generally nuclear dominated at forward angles due to a suppression of the Coulomb contribution resulting from the $N=Z$ cluster structure of $\prescript{6}{}{\rm Li}$. Moreover, we identify the important role that nuclear breakup and higher-order Coulomb effects play at low energies. These observations highlight the challenges of extracting capture cross sections from breakup data and the need to advance alternative indirect methods, such as electron-induced dissociation.

\section*{Acknowledgments}
This project was funded by the Deutsche Forschungsgemeinschaft (DFG, German Research Foundation) – project number: 514321794 (CRC1660: Hadron and Nuclei as discovery tools) and the PRISMA++ (Precision Physics, Fundamental Interactions and Structure of Matter) Cluster of Excellence.
\bibliography{references.bib} 

\end{document}